\documentclass[a4paper,french,11pt,fancyhdr]{article}

\usepackage[T1]{fontenc}
\usepackage[latin1]{inputenc}
\usepackage{babel}

\usepackage{amsmath}
\usepackage{amssymb}
\usepackage{mathrsfs}

\usepackage{color}
\usepackage{graphicx}
\graphicspath{{./images/}}
\usepackage{epsfig}
\usepackage{subfigure}
\usepackage{float}

\newcommand{\jpolHuge}[1]{\usefont{OT1}{pnc}{m}{n}
    \Huge\textsf{#1}
    \usefont{T1}{cmr}{m}{n}}

\newcommand{\jpolbf}[1]{\usefont{OT1}{pnc}{m}{n}
    \textbf{\textsf{#1}}
    \usefont{T1}{cmr}{m}{n}}

\newcommand{\etal}{\emph{et al.}}

\begin{document}

\noindent{\jpolHuge{Detuning effects on the mazer}}\\

\noindent J. Martin and T. Bastin\\
Institut de Physique Nucl\'eaire, Atomique et de Spectroscopie,\\
Université de Liège au Sart Tilman, B\^at. B15, B-4000 Liège, Belgium\\
\\
November 2003
\vspace{-135pt} \vspace{5cm}

\emph{A short review of the theoretical studies of the cold atom
micromaser (mazer) is presented. Existing models are then improved
by considering more general working conditions. Especially, the
mazer physics is investigated in the situation where a detuning
between the cavity mode and the atomic transition frequency is
present. Interesting new effects are pointed out. Especially, it
is shown that the cavity may slow down or speed up the atoms
according to the sign of the detuning and that the induced
emission process may be completely blocked by use of a positive
detuning. The transmission probability of ultracold atoms through
a micromaser is also studied and we generalize previous results
established in the resonant case. In particular, it is shown that
the velocity selection of cold atoms passing through the
micromaser can be very easily tuned and enhanced using a
nonresonant field inside the cavity. This manuscript is a summary
of Refs.~\cite{Mar02, Bas03b, Mar04}.} \vspace{10pt}

\section*{\jpolbf{Introduction}}

Laser cooling of atoms is a rapidly developing field in quantum
optics. Cold and ultracold atoms (temperature of the order of or
less than 1 $\mu$K) introduce new regimes in atomic physics often
not considered in the past. In particular, Englert
\etal~\cite{Eng91} have demonstrated new interesting properties in
the interaction of cold atoms with a micromaser field (see
Fig.~1). They have shown that excited atoms incident upon the
entrance port of a maser cavity will be reflected half of the
time, if the atoms are slow enough, even when the maser field is
in its vacuum state. This happens because the interaction
strength, between the atom and the maser field, changes strongly
when passing from the exterior to the interior of the cavity. More
recently, Scully \etal~\cite{Scu96} have shown that a new kind of
induced emission occurs when a micromaser is pumped by ultracold
atoms, requiring a quantum-mechanical treatment of the
center-of-mass motion. To insist on the importance of this
quantization usually defined along the $z$ axis, the system was
called mazer (for \underline{m}icrowave \underline{a}mplification
via \underline{$z$}-motion-induced \underline{e}mission of
\underline{r}adiation). The complete quantum theory of the mazer
has been first described in a series of three papers by Scully and
coworkers~\cite{Mey97,Lof97,Sch97}. The theory was written for
two-level atoms interacting with a single mode of a high-$Q$
cavity. In particular it was shown that the induced emission
properties are strongly dependent on the cavity mode profile.
Results were presented for the mesa, sech$^2$ and sinusoidal
modes. Retamal \etal~\cite{Ret98} later refined these results in
the special case of the sinusoidal mode, and a numerical method
was proposed by Bastin and Solano~\cite{Bas00} for efficiently
computing the mazer properties with arbitrary cavity field modes.
L\"offler \etal~\cite{Lof98} also demonstrated that the mazer can
act as a velocity selection device for an atomic beam and Bastin
and Solano~\cite{Bas03a} studied the trapping state properties of
the system. The mazer concept was extended by Zhang
\etal~\cite{Zha99, Zha98, Zha99b}, who considered two-photon
transitions~\cite{Zha99}, three-level atoms interacting with a
single cavity~\cite{Zha98} and with two cavities~\cite{Zha99b}.
Collapse and revival patterns with a mazer have been computed by
Du \etal~\cite{Du99}. Arun \etal~\cite{Aru00, Aru02} studied the
mazer with bimodal cavities and Agarwal and Arun~\cite{Aga00}
demonstrated resonant tunneling of cold atoms through two mazer
cavities.

\begin{figure}
\centering \epsfig{file=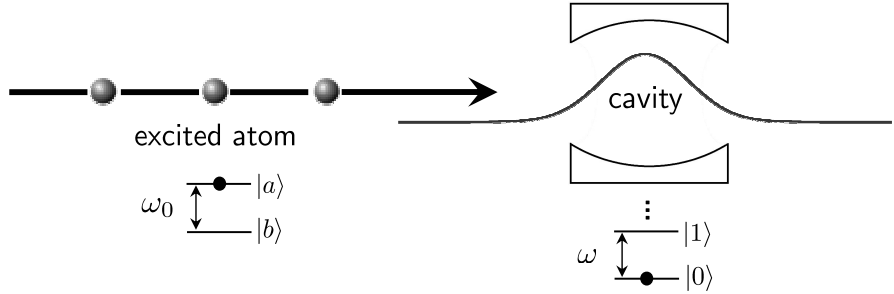, width=12cm} \caption{Micromaser
pumped by two-level atoms.}
\end{figure}

In all these previous studies, the mazer properties were always
presented in the resonant case where the cavity mode frequency
$\omega$ is equal to the atomic transition frequency $\omega_0$.
In this paper (see also Refs.~\cite{Mar02,Bas03b,Mar04}), we
remove this restriction and present the properties of the mazer in
the nonresonant case ($\omega \neq \omega_0$).

The paper is organized as follows. We first give a short review of
the maser action. Next we describe the mazer in the resonant case
($\omega=\omega_0$). We then discuss our results in the
nonresonant case ($\omega\ne\omega_0$). The photon emission
process inside the cavity and the transmission properties of the
mazer are especially investigated.

\section*{\jpolbf{Maser action}}

The interaction of a two-level atom with a single mode of an
electromagnetic field is well described by the Jaynes and Cummings
Hamiltonian~\cite{Jay63}
\begin{equation}
    \label{JCHamiltonian}
    \hat{H}_{JC} = \underbrace{\hbar \omega \hat{a}^{\dagger} \hat{a}}_{\hat{H}_{\textrm{field}}} + \underbrace{\hbar
    \omega_0 \hat{\sigma}^{\dagger} \hat{\sigma}}_{\hat{H}_{\textrm{atom}}} + \underbrace{\hbar g \big(
    \hat{\sigma} \hat{a}^{\dagger} + \hat{a}
    \hat{\sigma}^{\dagger}\big)}_{\hat{H}_{\textrm{interaction}}}
\end{equation}
where $\omega_0$ is the atomic transition frequency, $\omega$ the
cavity field mode frequency, $\hat{\sigma} = |b \rangle \langle
a|$ ($|a\rangle$ and $|b\rangle$ are respectively the upper and
lower levels of the two-level atom), $\hat{a}$ and
$\hat{a}^{\dagger}$ are respectively the annihilation and creation
operators of the cavity radiation field, and $g$ is the atom-field
coupling strength.

The behavior of this system is well-known. If the atom is
initially in the excited state $|a\rangle$ and the cavity field in
the Fock state $|n\rangle$, then the probability to find the atom
in the lower state $|b\rangle$ at a later time $t$ is given by
\begin{equation}
    \label{PabRabi}
    \mathcal{P}_{ab}(t) = \sin^2 2 \theta_n \sin^2 \frac{\sqrt{\delta^2 + \Omega_n^2} \,\, t}{2}
\end{equation}
where $\delta$ is the detuning $\omega$ - $\omega_0$, $\Omega_n$
the Rabi frequency $2 g \sqrt{n + 1}$ and $\theta_n$ the angle
defined by
\begin{equation}
    \cot 2\theta_n = -\frac{\delta}{\Omega_n}
\end{equation}

The atom carries out oscillations between the upper and the lower
energy states. This cycle of emission-absorption is called a Rabi
oscillation.

When thermal atoms, initially prepared in the excited state
$|a\rangle$, travel one by one through a maser cavity,
Eq.~(\ref{PabRabi}) gives the probability to find the atom at the
exit of the cavity in the state $|b\rangle$ provided $t$
represents the atom-field interaction time, that is $t = L / v$
with $L$ the cavity length and $v$ the velocity of the atom.

\section*{\jpolbf{Mazer action}}

If colder and colder (\textit{i.e.}\ slower and slower) atoms are
injected in the maser cavity, a quantized description of the
center-of-mass motion needs to be done as soon as the atomic
kinetic energy becomes of the order of or lower than the
interaction energy $\hbar g$. In this case, the Hamiltonian that
has to be considered reads
\begin{equation}
    \label{Hmazer}
    \hat{H} = \frac{\hat{p}_z^2}{2 m} + \hbar \omega \hat{a}^{\dagger} \hat{a} + \hbar
    \omega_0 \hat{\sigma}^{\dagger} \hat{\sigma} + \hbar g\,u(\hat{z}) \big(
    \hat{\sigma} \hat{a}^{\dagger} + \hat{a}
    \hat{\sigma}^{\dagger}\big)
\end{equation}
where $p_z$ is the atomic center-of-mass momentum along the $z$
axis, $m$ the atomic mass and $u(z)$ the cavity field mode
function modelling the spatial variations of the atom-field
interaction. The system described by the
Hamiltonian~(\ref{Hmazer}) is called \emph{mazer} \cite{Scu96}.

\subsection*{Resonant case : $\omega=\omega_0$}

In the resonant case, the system dynamics governed by the
Hamiltonian~(\ref{Hmazer}) may be easily studied in the atomic
dressed state basis $|\gamma^{\pm}_n\rangle =
\frac{1}{\sqrt{2}}\left( |a,n\rangle \pm |b,n+1\rangle\right)$. In
this basis, we show that the global wavefunction components
$\psi_n^{\pm}(z,t) = \langle z, \gamma^{\pm}_n| \Psi(t)\rangle$
(where $|\Psi(t)\rangle$ is the global wavefunction of the system)
obey a Schr\"odinger equation describing an elementary
one-dimensional scattering process upon the well defined
potentials $V^{\pm}_n =\pm \hbar g \sqrt{n + 1}\;u(z)$ (see
Ref.~\cite{Eng91}). If initially the incoming atoms upon the
cavity are in the excited state $|a\rangle$ and the field contains
$n$ photons, the atom-field state has initially two non-vanishing
components in the dressed state basis (the $\psi_n^{\pm}(z,t)$
components) and each of them is scattered differently. This
process results in a possible reflection or transmission of the
atoms by or through the cavity. These processes are accompanied by
a probability of finding the atom at the end in the lower state
$|b\rangle$, raising the photon number inside the cavity by one
unity (photon emission process). An analytical calculation of
these probabilities has been obtained in the particular cases
where the cavity field mode function is either given by the mesa
function ($u(z) = 1$ inside the cavity, $0$ elsewhere) either by
the sech$^2$ function which approximates the gaussian mode (see
Refs.~\cite{Mey97, Lof97}). An efficient numerical method has been
proposed by Bastin and Solano~\cite{Bas00} to compute the mazer
properties for an arbitrary field mode.

If we denote by $k$ the wave number of the incoming atom and by
$\kappa$ the particular wave number for which the atomic kinetic
energy $\hbar^2 \kappa^2/2m$ equals the vacuum coupling energy
$\hbar g$, we distinguish 3 regimes~: the hot atom regime ($k \gg
\kappa$), the intermediary regime ($k \simeq \kappa$) and the cold
atom regime ($k \ll \kappa$).

\begin{figure}[hbt]
\centering
\includegraphics*[width=11cm,bb=114 494 595 712,draft=false]{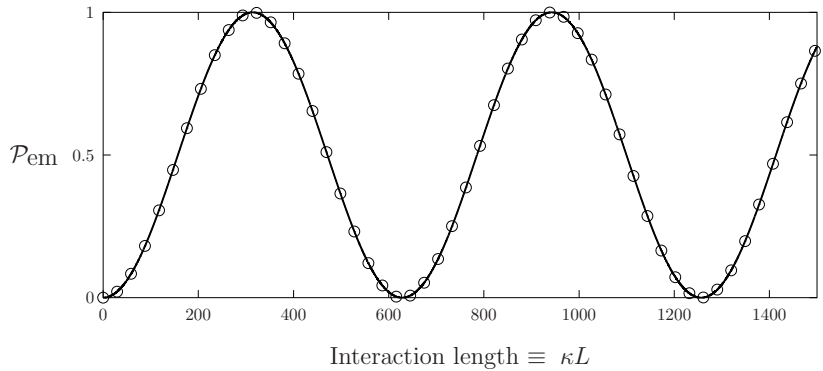}
\caption{Photon emission probability in the hot atom regime
($k/\kappa=100$). Plain curve : mazer Hamiltonian. Dots : J-C
Hamiltonian.}\label{rabioscillation}
\end{figure}

In the hot atom regime, the quantization of the atomic motion does
not play any role in the system dynamics (the scatterer potentials
are too small compared to the atomic kinetic energy). This is well
illustrated in Fig.~\ref{rabioscillation} which displays the
photon emission probability computed both from the mazer
Hamiltonian~(\ref{Hmazer}) and from the Jaynes-Cummings one.

\begin{figure}[H]
\centering
\includegraphics*[width=11cm,bb=114 492 596 718,draft=false]{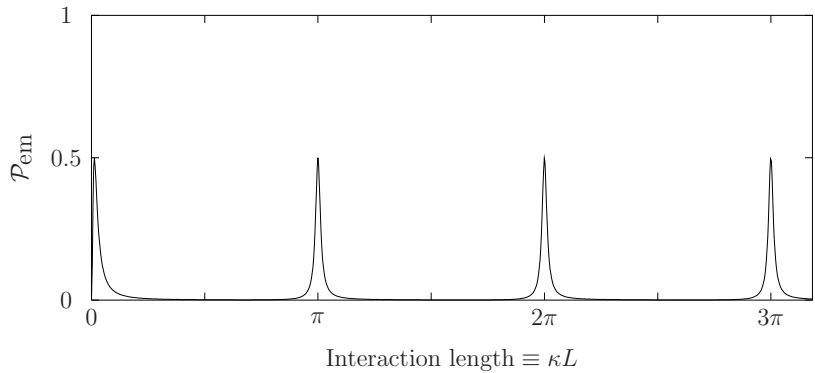}
\caption{Photon emission probability in the cold atom regime
($k/\kappa=0.01$) for a mesa mode function.}\label{pem-mazerh}
\end{figure}

Scully \etal~\cite{Mey97,Lof97,Sch97} remarkably showed that, in
the cold atom regime, the wave properties of the atoms become
important and the results predicted on the basis of the mazer
Hamiltonian (\ref{Hmazer}) completely differ from those predicted
with the Jaynes-Cummings one. In particular, the behavior of the
photon emission probability $\mathcal{P}_{\textrm{em}}$ changes
completely compared to the hot atom regime. For $k/\kappa\ll 1$
and the mesa mode function, $\mathcal{P}_{\textrm{em}}$ shows very
sharp resonances versus the interaction length $\kappa L$ as shown
in Fig.~\ref{pem-mazerh}.

\subsection*{Nonresonant case : $\omega\neq\omega_0$}

\begin{figure}
\centering
\includegraphics*[width=9cm,bb=131 95 663 390,clip=true]{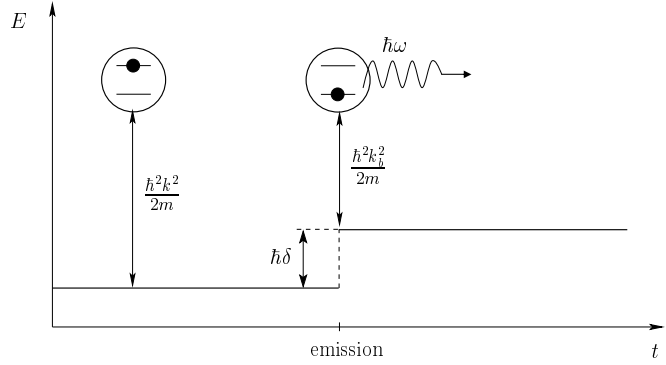}
\caption{Potential step effect of the cavity when a photon is
emitted by the atom. $E$ represents the total energy of the
atom-field system.}\label{step}
\end{figure}

In the nonresonant case, the equations verified by the
wavefunction components take a much more complicated form than in
the resonant case. It has however been possible to obtain
analytical expressions of the transmission and the photon emission
probabilities for a mesa mode function (see Ref.~\cite{Bas03b} for
more details). In this case, interesting new effects have been
obtained compared to the resonant case. Especially, the cavity may
slow down or speed up the atoms, according to the sign of the
detuning, and the induced emission probability may be completely
forbidden for positive detunings. These new effects are easily
understandable by considering the energy conservation depicted in
Fig.~\ref{step}. When, after leaving the cavity region, the atom
is passed from the excited state $|a\rangle$ to the lower state
$|b\rangle$, the photon number has increased by one unit in the
cavity and the internal energy of the atom-field system has varied
by the quantity $\hbar \omega - \hbar \omega_0 = \hbar \delta$.
This variation needs to be exactly counterbalanced by the external
energy of the system, i.e.\ the atomic kinetic energy. In this
sense, when a photon is emitted inside the cavity by the atom, the
cavity acts as a potential step $\hbar \delta$ (see
Fig.~\ref{step}). We denote by $k_b$ the wave number of the atom
after emission of a photon. The atomic transition $|a\rangle
\rightarrow |b\rangle$ induced by the cavity is therefore
responsible for a change of the atomic kinetic energy. According
to the sign of the detuning, the cavity will either speed up the
atom (for $\delta < 0$) or slow it down (for $\delta > 0$).

The use of positive detunings in the atom-field interaction
defines a well-controlled cooling mechanism. A single excitation
exchange between the atom and the field inside the cavity is
sufficient to cool the atom to a desired temperature $T = \hbar^2
k_b^2/2 m k_B$ ($k_B$ is the Boltzmann constant) which may be in
principle as low as imaginable. However, if the initial atomic
kinetic energy $\hbar^2 k^2/2m$ is lower than $\hbar \delta$ i.e.\
if $k/\kappa < \sqrt{\delta/g}$), the transition $|a,n\rangle
\rightarrow |b,n+1\rangle$ cannot take place (as it would remove
$\hbar \delta$ from the kinetic energy) and no photon can be
emitted inside the cavity. In this case the emission process is
completely blocked.

\begin{figure}[H]
\centering
\includegraphics*[width=11cm,bb=114 492 596 718,draft=false]{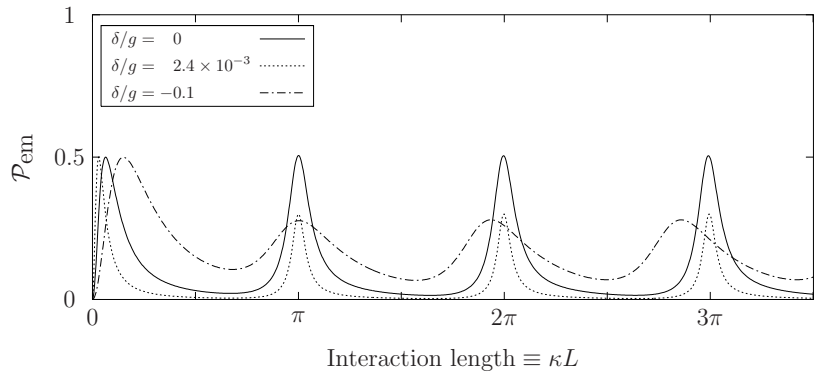}
\caption{Photon emission probability in the cold atom regime
($k/\kappa=0.05$) for different detuning values and in the case of
a mesa mode function.}\label{pemmazerdsg}
\end{figure}

Figure~\ref{pemmazerdsg} illustrates the induced emission
probability with respect to the interaction length $\kappa L$ for
various values of the detuning and in the cold atom regime. Like
the resonant case, the curves present a series of peaks where the
induced emission probability is optimum. However, the detuning
strongly affects the peak position, amplitude and width.

\begin{figure}[H]
\begin{center}
\noindent\mbox{\includegraphics[width=11cm, bb= 118 490 490 724,
clip = true]{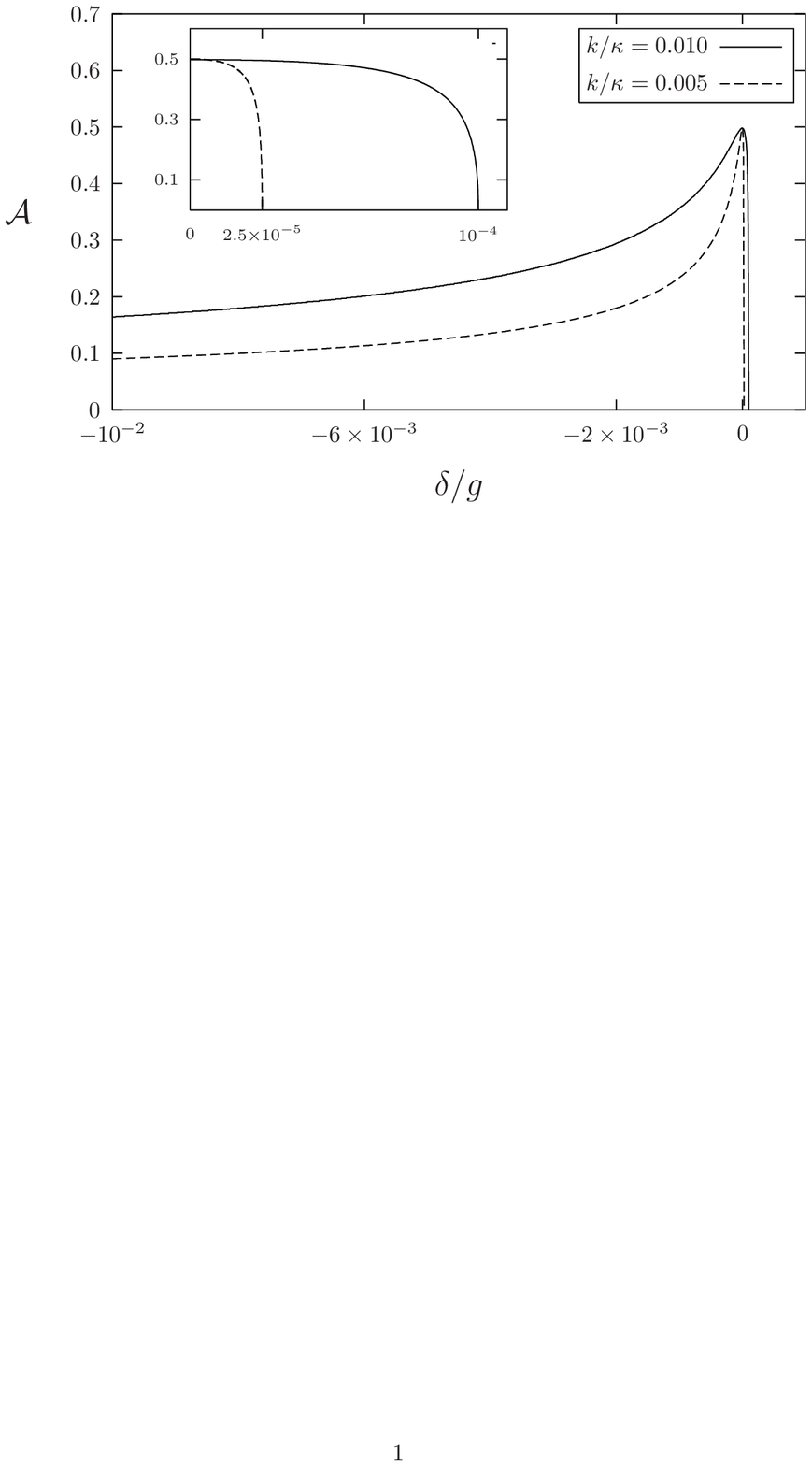}}
\end{center}
\vspace{-0.6cm} \caption{Amplitude $\mathcal{A}$ of the resonances
with respect to $\delta/g$ for 2 values of $k/\kappa$ in the cold
atom regime.} \label{PeakAmpFig}
\end{figure}

We illustrate in Fig.~\ref{PeakAmpFig} the amplitude $\mathcal{A}$
of the resonance peaks as a function of the detuning. Contrary to
the hot atom regime where the amplitude of the Rabi oscillations
is given by the factor $\sin^2 2 \theta_n$ (see
Eq.~(\ref{PabRabi})) which is insensitive to the sign of the
detuning, the curves in Fig.~\ref{PeakAmpFig} present a strong
asymmetry with respect to the sign of $\delta$. This results from
the potential step $\hbar \delta$ felt by the atoms when they emit
a photon. For cold atoms whose energy is similar or less than the
step height, the sign of the step is a crucial parameter. The
induced emission probability drops very rapidly down to zero for
positive detunings, in contrast to what happens for negative
detunings. It is very interesting to note that the peak amplitude
is equal to the amplitude at resonance ($1/2$) times the
transmission factor of a particle of momentum $\hbar k$ through a
potential step $\hbar\delta\,$ ($(4 k_b/k)/(1 + k_b/k)^2$). This
is an additional argument to say that the use of a detuning adds a
potential step effect for the atoms emitting a photon inside the
cavity (see Fig.~\ref{step}).

\section*{\jpolbf{Transmission properties of the mazer}}

L\"{o}ffler \etal~\cite{Lof98} have proposed recently to use the
mazer for narrowing the velocity distribution of an ultracold
atomic beam. This could be very useful to define long coherence
lengths. We investigated the effects of a detuning on the
transmission probability of an atom through the mazer and on the
velocity selection process (see Ref.~\cite{Mar04}). We found that
the atomic transmission probability through the cavity shows with
respect to the detuning fine resonances that could be very useful
to define extremely accurate atomic clocks. It also turns out that
the velocity selection in an atomic beam could be significantly
enhanced and easily tuned by use of a positive detuning.

\subsection*{Transmission probability}

\begin{figure}[H]
\centering
\includegraphics*[width=10cm,bb=111 493 484 714,draft=false]{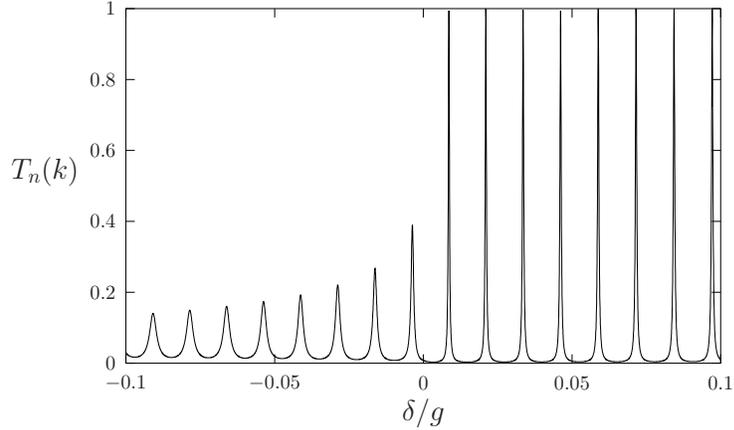}
\caption{Transmission probability of an excited atom through the
mazer with respect to the detuning ($k/\kappa = 0.05$, $\kappa L =
1000$, $n = 0$)} \label{TdsgFig}
\end{figure}

Figure~\ref{TdsgFig} illustrates the atomic transmission
probability with respect to the detuning (in the ultracold
regime). The curve obtained presents very sharp resonances. For
realistic experimental parameters (see discussion
in~\cite{Lof97}), these resonances may even become extremely
narrow. Their width amounts only $10^{-2}$~Hz for $\kappa L =
10^5$, $g = 100$ kHz and $k/\kappa = 0.01$. This could define very
useful metrology devices (atomic clocks for example) based on a
single cavity passage and with better performances than what is
usually obtained in the well known Ramsey configuration with two
cavities or two passages through the same cavity~\cite{Cla91}.

\subsection*{Velocity selection}

\begin{figure}[H]
\includegraphics*[width=8cm, bb=115 285 485 535,clip=true]{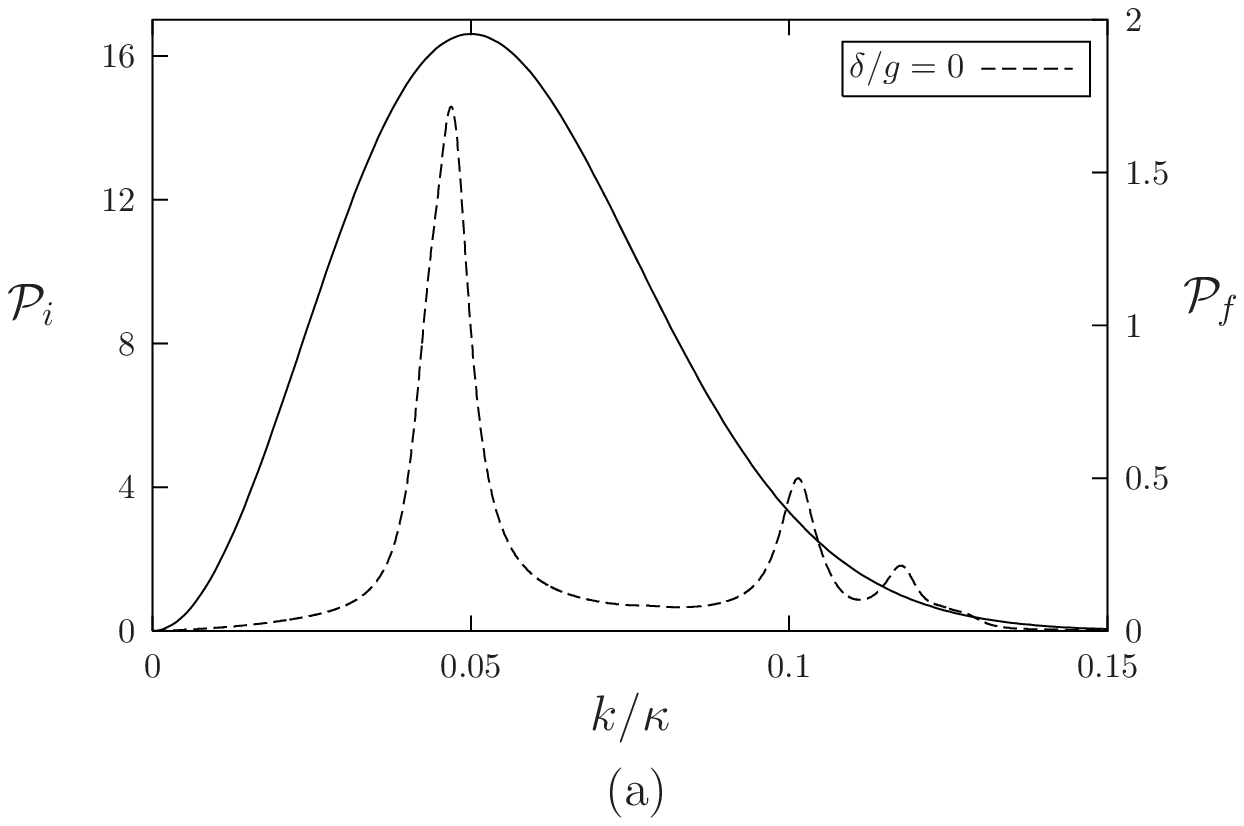}
\includegraphics*[width=8cm, bb=115 285 485 535,clip=true]{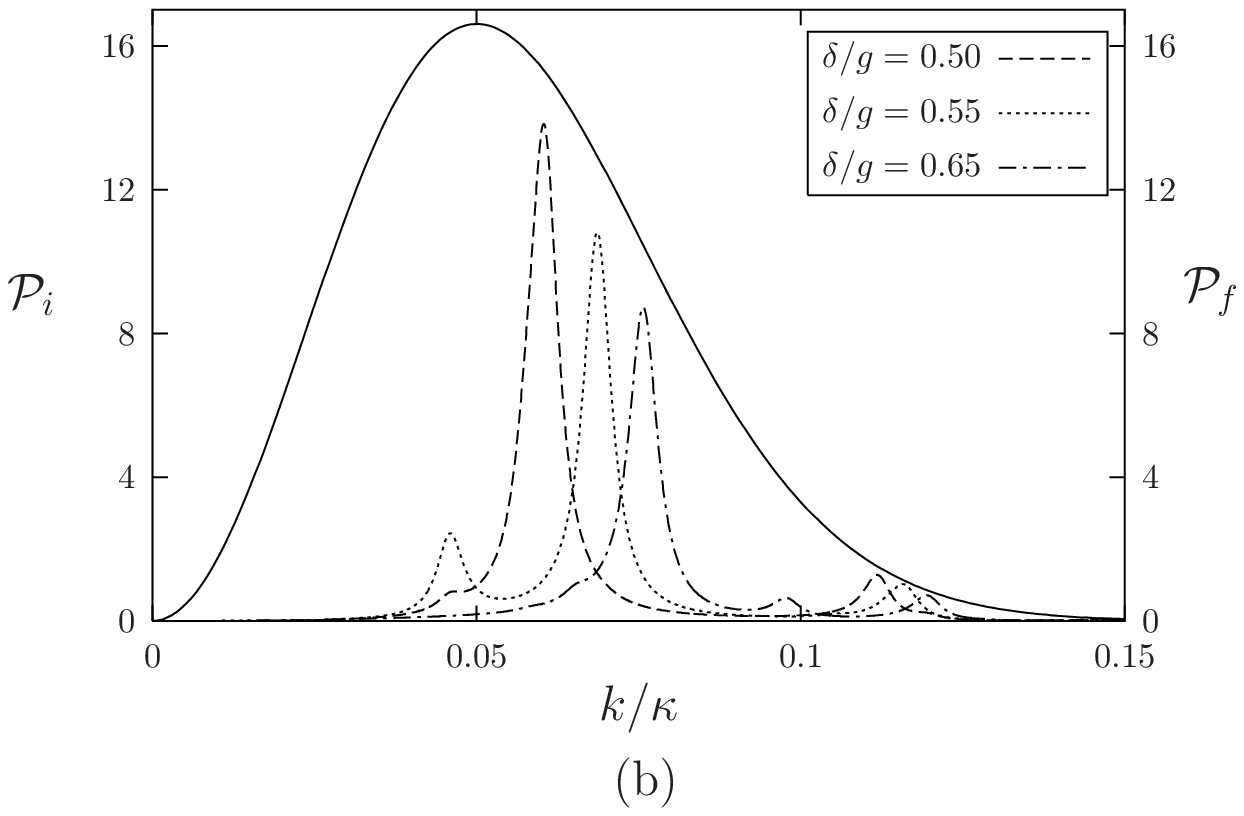}
\caption{Initial (plain curve) and final (dashed curves) velocity
distributions (a) at resonance and (b) for various detuning
values.} \label{PifFigs}
\end{figure}

We show in Figs.~\ref{PifFigs} how a Maxwell-Boltzmann
distribution (with $k_0/\kappa = 0.05$ where $k_0$ is the most
probable wave number) is affected when the atoms are sent through
the cavity. The cavity parameters have been taken identical to
those considered in \cite{Lof98} to underline the detuning
effects. We see from these figures that the final distributions
are dominated by a narrow single peak whose position depends
significantly on the detuning value. This could define a very
convenient way to select any desired velocity from an initial
broad distribution. Also, notice from the $\mathcal{P}_f$ scale
that a positive detuning significantly enhances the selection
process.

\renewcommand{\refname}{\jpolbf{References}}

\end{document}